\documentclass[10pt]{iopart}

\usepackage[dvips]{graphics}
\usepackage{graphicx}
\usepackage{float}
\usepackage{appendix}
\usepackage{times}

\def\iomn{i\omega_n}
\def\tn{\widetilde{n}}
\def\tp{\widetilde{p}}
\def\tphi{\widetilde{\varphi}}
\def\tPsi{\widetilde{\Psi}}
\def\bra{\langle}
\def\ket{\rangle}

\def\vk{{\bf k}}
\def\vr{{\bf r}}

\def\bR{{\bf R}}

\def\vR{{\bf R}}
\def\Tr{{\rm Tr}}

\newcommand{\bk}{\mathbf{k}}

\newcommand{\tchikm}{|\tilde{\chi}^{\bR}_{\bk m}\rangle}

\newcommand{\wfkm}{|w^{\bR}_{\bk m}\rangle}

\makeatletter
\newlength{\earraycolsep}
\def\eqnarray{\stepcounter{equation}\let\@currentlabel%
\theequation
\global\@eqnswtrue\m@th
\global\@eqcnt\z@\tabskip\@centering\let\\\@eqncr
$$\halign to\displaywidth\bgroup\@eqnsel\hskip\@centering
$\displaystyle\tabskip\z@{##}$&\global\@eqcnt\@ne
\hskip 2\earraycolsep \hfil$\displaystyle{##}$\hfil
&\global\@eqcnt\tw@ \hskip 2\earraycolsep
$\displaystyle\tabskip\z@{##}$\hfil
\tabskip\@centering&\llap{##}\tabskip\z@\cr}
\makeatother

\def\ee{\end{equation}}
\def\theequation{\thesection.\arabic{equation}}

\makeatletter
\renewcommand\theequation{\thesection.\arabic{equation}}
\makeatother
\def\bra{\langle}
\def\ket{\rangle}

\def\vk{{\bf k}}
\def\vr{{\bf r}}
\def\vR{{\bf R}}
\begin{document}

\title[A Self-consistent DFT+DMFT scheme in the Projector Augmented Wave]{A Self-consistent DFT+DMFT scheme in the Projector Augmented Wave :  Applications  to Cerium, Ce$_2$O$_3$ and Pu$_2$O$_3$
with the Hubbard I solver and comparison to DFT+U.}

\author{B.~Amadon}
\address{CEA, DAM, DIF, F 91297 Arpajon, France}
\ead{bernard.amadon@cea.fr}
\begin{abstract}
An implementation of the full self-consistency over electronic density
in the DFT+DMFT framework on  the basis of 
a plane wave-projector augmented wave (PAW) DFT code is presented.
It allows for an accurate calculation of total energy in DFT+DMFT within a plane wave approach.
Contrary to frameworks based on maximally localized Wannier function, the method easily apply to $f$ electron systems, such as Cerium, Cerium oxide Ce$_2$O$_3$ and Plutonium oxide Pu$_2$O$_3$. 
In order to have a correct and physical calculation of the energy terms, we find that
the calculation of the self-consistent density is mandatory.
The formalism is general and does not depend on any method used to solve the impurity model.
Calculations are carried out within the Hubbard I approximation, which is fast to solve, and gives a good
description of strongly correlated insulators. 
We compare DFT+DMFT and DFT+U solutions, and underline the qualitative differences of their converged densities.
We emphasize that contrary to DFT+U, DFT+DMFT does not break the spin and orbital symmetry.
As a consequence, DFT+DMFT implies, on top of a better physical description  of correlated metals and insulators, a reduced
occurrence of unphysical metastable solutions in correlated insulators in comparison to DFT+U.

\end{abstract}

\maketitle

\section{Introduction}
Electrons in localized orbitals exhibit strong interaction effects.
Their ab-initio description, in condensed matter physics, have made progress
thanks to the development of the DFT+U\cite{Anisimov91,Anisimov97} and 
DFT+DMFT methods\cite{georges96,Anisimov97,Lichtenstein98,georges2004,Kotliar06,Held07}.
This last method has in particular been very successful
to describe both the itinerant and localized behaviour of strongly interacting electrons.

Whereas DFT+U is implemented routinely in DFT electronic structure codes to compute
all kinds of properties, frameworks to implement DFT+DMFT for an arbitrary basis
(e.g plane waves) have emerged only recently.
Two different schemes both using Wannier functions were used:
the construction\cite{marzari_wannier_1997_prb} 
of Maximally Localized Wannier Function (MLWF)\cite{Lechermann06} or NMTO\cite{Pavarini04}
and the calculation of Projected Local Orbitals (PLO)\cite{Anisimov05,Amadon08,Korotin08,Aichhorn09,Haule10}.
The PLO framework is light and does not require the calculation
of MLWF: in particular, MLWF for f-electrons systems might be technically difficult
to compute.

Besides, the need for a precise method to compute total energy requires 
the self-consistency over the electronic density
to be performed \cite{savrasov_kotliar_pu_nature_2001,min05,Pourovskii07,dimarco09,Haule10,Suzuki10,Shick09}.
Calculations of total energy using the self-consistency over density
are scarce in DFT+DMFT\cite{savrasov_kotliar_pu_nature_2001,Pourovskii07,dimarco09,Shick09,Haule10}.
Simplified schemes designed for the Hubbard I approximation also included the self-consistency
and were successful in describing actinides compounds\cite{Shick09}.

In this paper, we present an implementation of DFT+DMFT  with total energy
in the PLO scheme, using the self-consistency over density in a plane wave (PW) based PAW code.
This implementation is done in the code ABINIT\cite{Abinit09,abinit3}
and is an extension of a previously published non self-consistent (NSCF) projected local orbital scheme
in PAW\cite{Amadon08}. Since then, a similar NSCF implementation from the output of the PAW code VASP\cite{vasp2} has been described\cite{Karolak10}.
The Projector Augmented Wave\cite{Blochl94} framework in combination with e.g a plane
wave basis is a widely used tool\cite{Kresse99,Holzwarth97,Torrent08} to carry out accurate electronic structure calculations.
The framework of our implementation is general and 
could also be used in e.g real-space based PAW code\cite{Enkovaara10,Ono10}.
Moreover, our implementation put the computation of energy in
DFT+DMFT on the same footing as LDA : a wide range of systems are available for description.

The plan of the paper is the following.
In Sec.\ref{sec:theo}, the PLO framework for DMFT is  briefly described, expressions for total energy are given and
the way the self-consistency is performed is detailed. We emphasize in particular that having two expressions
for the total energy is a check to avoid any errors in the implementation. Technicalities
related to the PAW formalism are given in \ref{app:scpaw} and \ref{app:etot_paw} for self-consistency and 
\ref{app:compaldau} for the PLO scheme.

In Sec.\ref{sec:appl}, we study three correlated systems with $f$ orbitals. 
Actinides and Lanthanides have recently attracted a lot of theoretical interest\cite{Larson07,Pourovskii07,Pourovskii09,Dorado09,Jiang09,Suzuki09,Petit10,Yin11}
For these systems, our PLO scheme for DFT+DMFT is particularly adapted: using MLWF would
be possible, but the determination of MLWF is presently not a straightforward task
for $f$ electrons systems.
We first investigate cerium and cerium oxide Ce$_2$O$_3$. 
A previous implementation of self-consistency in DFT+DMFT using an ASA basis was applied to these compounds \cite{Pourovskii07}.
This is thus the opportunity to put forward the improvement brought by our approach
on these systems: because of the PW-PAW basis, the physical accuracy is better, and the method is more flexible.
Besides, we illustrate, for the sake of clarity, the difference between self-consistent densities obtained
in DFT+U and DFT+DMFT for Ce$_2$O$_3$.
We then present a study 
of Pu$_2$O$_3$. We compare this compound to cerium oxide and our results
to previous DFT+U calculations.
For these systems, we emphasize that DFT+DMFT should, decrease the number of spurious metastables states
as found in DFT+U\cite{Shick00,Larson07,Amadon08a,Jomard08,Dorado09,Jollet09,Meredig10}.
Especially for cerium,  on the basis of our calculations on $\gamma$ cerium,
the calculation of complex phases of cerium, such as $\beta$ cerium, should be easier in comparison
to DFT+U\cite{Amadon08a}.
In this view, simple Hubbard I approximation would be useful for large scale calculations on defects and surfaces.

In \ref{app:compaldau}, we show that our implementation can be precisely compared
with an DFT+U implementation, when the self-energy in DMFT
is chosen to be the static mean field
self-energy.  We illustrate this on the case of cerium.
The agreement between the two calculations
shows the consistency of the two implementations.
In \ref{app:bands}, we discuss the impact of the windows of energy
used to compute Wannier orbitals on the spectral function of cerium: it underlines
the need for a consistent determination of U with the choice of the energy window.

\section{Theoretical Framework}
\label{sec:theo}

\subsection{DFT+DMFT formalism: projected local orbitals}
In DFT+U or in the combination of Density Functional Theory and Dynamical Mean Field Theory,
a peculiar treatment is reserved for {\it correlated electrons}. As a consequence, 
{\it correlated orbitals} need to be defined: this is the focus of this subsection.

We first summarize the main equations from the Projected local orbitals framework
\cite{Anisimov05,Lechermann06,Amadon08,Korotin08}.  
In this framework, a subspace $\mathcal{W}$ of the total
Hilbert space is used as a basis for the DMFT calculation. 
This subspace is spanned by a given number of DFT Kohn Sham (KS) Bloch wave functions.
The local orbital on which the coulomb interaction applies are
defined on the following way: 
The KS Bloch wave function $\Psi_{\bk\nu}$ are projected on local 
orbitals $\chi^{\bR}_{\bk m}$ (e.g. atomic orbitals). 
Let's define $P^{\bR}_{m\nu}(\bk)$ such as:
\begin{equation}\label{eq:pro_chipsi}
P^{\bR}_{m\nu}(\bk)\equiv\langle\chi^{\bR}_{\bk m}|\Psi_{\bk\nu}\rangle,
\end{equation}
where m = 1 , . . . , M is an orbital index within
the correlated subset, $\bR$ denotes the correlated atom within the primitive
unit cell, $\bk$ is a k-point in the Brillouin Zone, and $\nu$ is the band index.
In \ref{subapp:localorb}, we discuss the technical calculation of the projection
in PAW.  This projection defines Wannier functions. These Wannier
functions are built upon the local
orbitals and are an orthonormalized linear combination of a limited number
of KS wave functions.
Let's define:
\begin{equation}
\tchikm\equiv\sum_{\nu\in {\cal W}}\langle\Psi_{\bk\nu}|\chi^{\bR}_{\bk m}\rangle
|\Psi_{\bk\nu}\rangle.
\label{eq:tchi}
\end{equation}
The orthonormalization of $\tchikm$ gives the Wannier functions $\wfkm$.
We call $\bar{P}^{\bR}_{m\nu}(\bk)$ the projections accordingly orthonormalized.

The flexibility in the choice of  $\mathcal{W}$, and thus in the choice
of Wannier functions enable us an easy comparison to calculations
using Maximally Localized Wannier Functions\cite{Lechermann06,Amadon08}
or atomic orbitals (which are used in the PAW DFT+U implementations \cite{Bengone00, Amadon08a}).

From the orthonormalized projections, the local quantities
can be linked to the quantity defined on the lattice. For the Green's function and the self-energy, 
we thus have:
\begin{eqnarray}
\nonumber
G^{\bR,{\rm imp}}_{mm'}(\iomn)\hspace{-0.1cm} &=&\hspace{-0.5cm}
\sum_{\bk,(\nu\nu')\in {\cal W}}
\bar{P}^{\bR}_{m\nu}(\bk)\,G^{\rm bl}_{\nu\nu'}(\bk,\iomn)\,
\bar{P}^{\bR*}_{\nu' m'}(\bk)\,,\,\,\label{eq:g_limband}\\
\\
\nonumber
\Delta\Sigma^{\rm bl}_{\nu\nu'}(\bk,\iomn)&=&
\sum_{\bR}\sum_{mm'}\bar{P}^{\bR*}_{\nu m}(\bk) \,\Delta\Sigma^{\rm imp}_{mm'}(\iomn)
\,\bar{P}^{\bR}_{m'\nu'}(\bk)\,,\\
\label{eq:Sband}
\end{eqnarray}
where
\begin{eqnarray}
\nonumber
\label{eq:Gband}
&&G^{\rm bl}_{\nu\nu'}(\bk,\iomn)=\left\{\left
[(\iomn+\mu-\epsilon_{\nu,\bk})\delta_{\nu_1,\nu'_1}-
\Delta\Sigma_{\nu_1,\nu'_1}^{\rm bl}(\bk,\iomn)\right]^{-1}\right\}_{\nu\nu'} \\
\\
&&\Delta\Sigma^{\rm imp}_{mm'}(\iomn)=\Sigma_{mm'}^{\rm{imp}}(\iomn)-
\Sigma_{mm'}^{\rm{dc}}\quad.
\end{eqnarray}
Eq. \ref{eq:Gband} contains an inversion of a band matrix (of dimension the size $\mathcal{W}$) for each frequency and k-point.
$\epsilon_{\nu,\bk}$ is the DFT Kohn Sham eigenvalue for band index $\nu$ and k-point $\bk$.
$\Sigma_{\rm imp}$ and $\Sigma_{\rm dc}$ are respectively the self-energy of the impurity problem,
and the double counting term. 
The self-energy of the impurity problem is computed with the Dyson Equation and is used to compute the lattice Green's function as expressed by equations \ref{eq:Gband} and \ref{eq:Sband}. The DMFT scheme is summarized on Fig \ref{fig:scfdmft}.

\subsection{DFT+DMFT formalism: self-consistency over density}

\begin{figure}
{\scalebox{0.4}{

\begin{picture}(0,0)%
\includegraphics{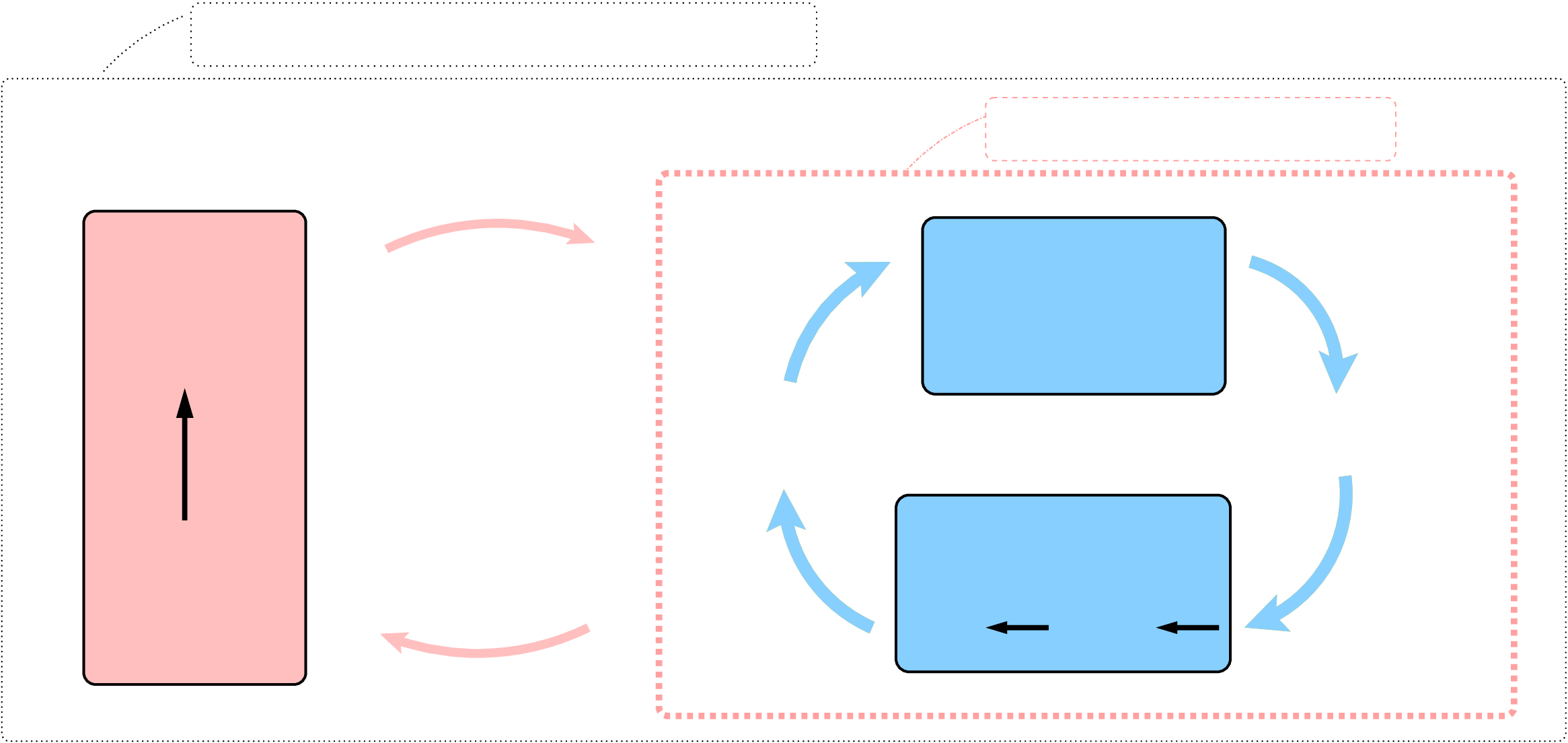}%
\end{picture}%
\setlength{\unitlength}{3947sp}%
\begingroup\makeatletter\ifx\SetFigFont\undefined%
\gdef\SetFigFont#1#2#3#4#5{%
  \normalfont\fontsize{#1}{#2pt}%
  \fontfamily{#3}\fontseries{#4}\fontshape{#5}%
  \selectfont}%
\fi\endgroup%
\begin{picture}(18644,8819)(-7446,-7733)
\put(7051,-4336){\makebox(0,0)[lb]{\smash{{\SetFigFont{44}{52.8}{\rmdefault}{\mddefault}{\updefault}$\Sigma=\mathcal{G}_0^{-1}-G^{-1}_{\rm imp}$}}}}
\put(676,-4336){\makebox(0,0)[lb]{\smash{{\SetFigFont{44}{52.8}{\rmdefault}{\mddefault}{\updefault}$\mathcal{G}_0^{-1}=\Sigma+G^{-1}_{\rm imp}$}}}}
\put(1351,-6211){\makebox(0,0)[lb]{\smash{{\SetFigFont{44}{52.8}{\rmdefault}{\mddefault}{\updefault}$G_{\rm imp}$}}}}
\put(-5549,-5986){\makebox(0,0)[lb]{\smash{{\SetFigFont{34}{40.8}{\rmdefault}{\mddefault}{\updefault}{$n(\bf r)$}%
}}}}
\put(8626,-6061){\makebox(0,0)[lb]{\smash{{\SetFigFont{44}{52.8}{\rmdefault}{\mddefault}{\updefault}$\Sigma$}}}}
\put(5176,-6436){\makebox(0,0)[lb]{\smash{{\SetFigFont{44}{52.8}{\rmdefault}{\mddefault}{\updefault}{$G_{\rm latt}$}%
}}}}
\put(8476,-2461){\makebox(0,0)[lb]{\smash{{\SetFigFont{44}{52.8}{\rmdefault}{\mddefault}{\itdefault}$G$}}}}
\put(-6074,-2686){\makebox(0,0)[lb]{\smash{{\SetFigFont{29}{34.8}{\rmdefault}{\mddefault}{\updefault}{$H_{\rm KS-DFT}$}%
}}}}
\put(4301,-3061){\makebox(0,0)[lb]{\smash{{\SetFigFont{29}{34.8}{\rmdefault}{\mddefault}{\updefault}{problem}%
}}}}
\put(4201,-2311){\makebox(0,0)[lb]{\smash{{\SetFigFont{29}{34.8}{\rmdefault}{\mddefault}{\updefault}{Impurity}%
}}}}
\put(3376,-5311){\makebox(0,0)[lb]{\smash{{\SetFigFont{29}{34.8}{\rmdefault}{\mddefault}{\updefault}{Self-consistency}%
}}}}
\put(5176,-5836){\makebox(0,0)[lb]{\smash{{\SetFigFont{29}{34.8}{\rmdefault}{\mddefault}{\updefault}{condition}%
}}}}
\put(-4799,464){\makebox(0,0)[lb]{\smash{{\SetFigFont{29}{34.8}{\rmdefault}{\mddefault}{\updefault}{DFT+DMFT Loop over density}%
}}}}
\put(-2849,-1111){\makebox(0,0)[lb]{\smash{{\SetFigFont{44}{52.8}{\rmdefault}{\mddefault}{\updefault}$\langle \chi^{\bf R}_{{\bf k}m}|\Psi_{{\bf k}\nu} \rangle $}}}}
\put(-2624,-6136){\makebox(0,0)[lb]{\smash{{\SetFigFont{44}{52.8}{\rmdefault}{\mddefault}{\updefault}$f_{\nu,\nu',{\bf k}}^{\rm DFT+DMFT}$}}}}
\put(4426,-586){\makebox(0,0)[lb]{\smash{{\SetFigFont{29}{34.8}{\rmdefault}{\mddefault}{\updefault}{DMFT for fixed $n({\bf r})$}%
}}}}
\put(-2474,-2536){\makebox(0,0)[lb]{\smash{{\SetFigFont{44}{52.8}{\rmdefault}{\mddefault}{\updefault}$\epsilon_{\nu,{\bf k}}^{\rm KS-DFT}$}}}}
\put(1276,-2611){\makebox(0,0)[lb]{\smash{{\SetFigFont{43}{51.6}{\rmdefault}{\mddefault}{\itdefault}{$\mathcal{G}_0$}%
}}}}
\put(3376,-6436){\makebox(0,0)[lb]{\smash{{\SetFigFont{44}{52.8}{\rmdefault}{\mddefault}{\updefault}{$G_{\rm imp}$}%
}}}}
\end{picture}%

}}
\centering
\caption{Fully Self-consistent DFT+DMFT scheme adapted to the Projected
Local Orbital scheme used in our implementation.
For a fixed electronic density, the DMFT Loop is represented in blue.
It contains two steps: Firstly, the impurity model is solved to compute
the Green's function $G$ and the self-energy $\Sigma$.
Secondly the lattice Green's function $G_{\rm latt}$ is computed from the impurity self-energy, and the self-consistency
condition states that the local Green's function is also the impurity Green's function.
When this DMFT loop is converged, one can compute 
the lattice Green's function and the
non diagonal occupations in the Kohn Sham Bloch basis.
The occupations are used to compute the total 
electronic density (in practice, its PW and PAW components) thanks
to Eq. \ref{eq:rhoscf}.
From this density, the Kohn Sham DFT Hamiltonian is built and diagonalized.
The new KS bloch wave functions, and the eigenvalues are then used to compute
the Green's function (Eq. \ref{eq:Gband}). Then a new DMFT loop is performed.
This cycle (in pink) is repeated until convergence of the density.}
\label{fig:scfdmft}
\end{figure}
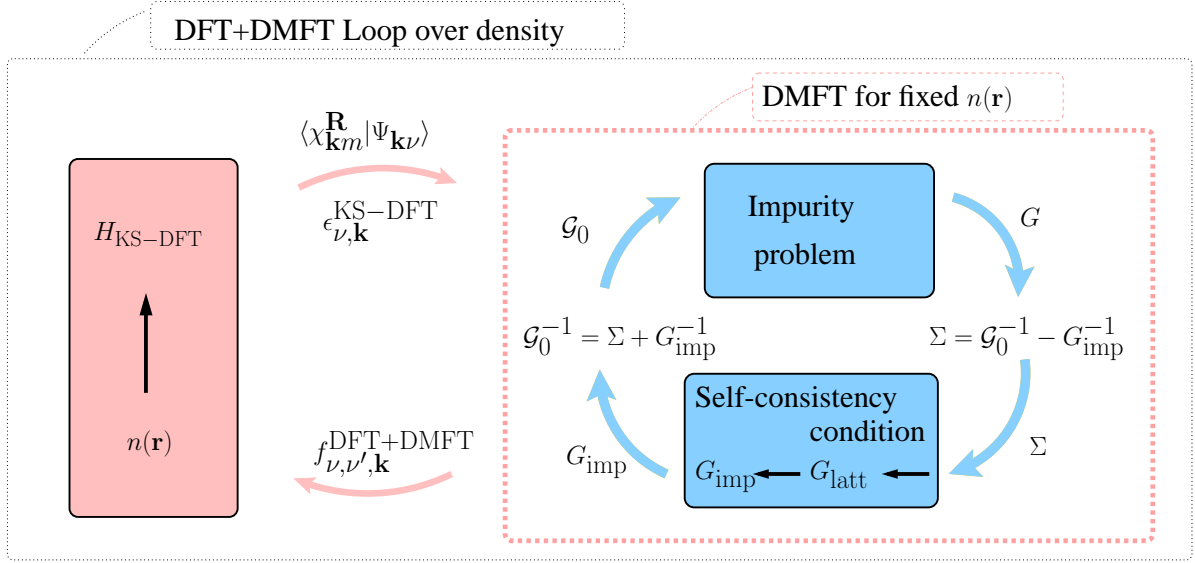

The scheme of the self-consistency over electronic density
within DFT+DMFT is shown on Fig. \ref{fig:scfdmft} and briefly described in the caption.
In more details, from the DMFT loop, the non-diagonal occupation
in the Bloch band index can be extracted from \ref{eq:Gband}
and used to compute the density using:
\begin{eqnarray}
\nonumber
n(\vr)&=&\bra \vr | \hat{n} | \vr \ket
=\sum_{\vk,\nu,\nu'}\bra \vr |\Psi_{\bk\nu}\ket \bra \Psi_{\bk\nu} | \hat{n} | 
\Psi_{\bk\nu'} \ket \bra \Psi_{\bk\nu'}| \vr \ket \\
&\equiv& 
\sum_{\vk,\nu,\nu'}\bra \vr |\Psi_{\bk\nu}\ket  
f_{\nu,\nu',\vk} \bra \Psi_{\bk\nu} |\vr \ket 
\label{eq:rhoscf}
\end{eqnarray}
$\hat{n}$ is the density operator in these equations.

Then, the DFT  Hamiltonian is built and diagonalized: new KS eigenvalues
and eigenfunctions are extracted.
A special care is taken to obtain the new KS bands with a good
accuracy for each new electronic density, especially for unoccupied KS states.
Then, projections \ref{eq:pro_chipsi} are recomputed
to build Wannier functions for the next DMFT loop. KS eigenvalues
are also used to compute the Green's function using \ref{eq:Gband}.

Peculiarities of the PAW formalism for the computation of the electronic density are described in \ref{app:scpaw}.

\subsection{Calculation of Internal Energy in DFT+DMFT}

The DFT+DMFT formalism can be derived from
a functional \cite{Savrasov04} of both the local density $n(\vr)$ and the local Green's function
$G^{\rm loc}$ \cite{Gloc}.
The internal energy can be derived and one obtain the general formula of Eq. 3 of Ref. \cite{Amadon06}.
We first write two terms which appears in this equation:
\begin{eqnarray}
\Tr[G_{\rm DFT}H_{\rm KS-DFT}]&=&\sum_{\nu,vk} f^{\rm DFT}_{\nu,\vk}[n({\vr})] \epsilon^{\rm KS-DFT}_{\nu,\vk}[n({\vr})] \\
\Tr[G_{\rm DFT+DMFT}H_{\rm KS-DFT}]&=&\sum_{\nu,vk} f^{\rm DFT+DMFT}_{\nu,\vk}[n({\vr}),G^{\rm loc}] \epsilon^{\rm KS-DFT}_{\nu,\vk}[n({\vr})]
\end{eqnarray}
In the above equations and below, $n(\vr)$ and $G^{\rm loc}$ are the converged density and local Green's function
of the full self-consistent DFT+DMFT cycle.
For a given electronic density $n(\vr)$, and thus a given Kohn Sham DFT Hamiltonian $H_{\rm KS-DFT}[n(\vr)]$,
$\epsilon^{\rm KS-DFT}_{\nu,\vk}[n(\vr)]$ are the Kohn Sham eigenvalues
of this Hamiltonian.
$G_{\rm DFT}$ is the  Kohn Sham DFT Green's function and $G_{\rm DFT+DMFT}$ is the DFT+DMFT Green's function. 
We stress that $G_{\rm DFT}$ and $G_{\rm DFT+DMFT}$ can both be computed on the basis of Kohn Sham wavefunctions.
But this basis is just a practical way to express these quantities.
Physically, $G_{\rm DFT+DMFT}$ is computed with Eq \ref{eq:Gband} which contains the current Kohn Sham Hamiltonian and the self-energy,
whereas  $G_{\rm DFT}$ is computed with the same expression but without the self-energy.
$f^{\rm DFT+DMFT}_{\nu,\vk}[n(\vr),G^{\rm loc}]$ is the DFT+DMFT occupations computed directly by integration
over frequency of the DFT+DMFT Green's function $G^{\rm DFT+DMFT}_{\nu}(\bk,\iomn)$ expressed in the Bloch Kohn Sham basis.
$f^{\rm DFT}_{\nu,\vk}[n(\vr)]$ is the Fermi Dirac occupations and 
can also be computed directly by integration
over frequency of the Kohn Sham DFT Green's function expressed in the Bloch Kohn Sham basis $G^{\rm DFT}_{\nu}(\bk,\iomn)$.

With the notation from Eq. 3 of Ref. \cite{Amadon06}, we have
$\Tr[G_{\rm DFT}H_{\rm KS-DFT}]=\sum_\lambda \epsilon_{\lambda}^{\rm KS}$
and $\Tr[G_{\rm DFT+DMFT}H_{\rm KS-DFT}]=\langle H_{\rm KS} \rangle$.

We can thus rewrite Eq. 3 of Ref. \cite{Amadon06}:

\begin{eqnarray}
\nonumber
E_{\rm DFT+DMFT}=\overbrace{E_{\rm DFT}[n({\vr})] 
		  -\Tr[H_{\rm KS-DFT}G_{\rm DFT}]}^{-E_{\rm Ha}[n(\vr)]+E_{\rm xc}[n(\vr)]-\int v_{\rm xc} n(\vr) d\vr} \\
		  +\Tr[H_{\rm KS-DFT}G_{\rm DFT+DMFT}]
\nonumber
		  + \bra H_U \ket -E_{\rm DC} \\
\label{eq:edftdmft}
\end{eqnarray}
In this equation, $E_{\rm Ha}[n(\vr)]$ and $E_{\rm xc}[n(\vr)]$ are the Hartree and the exchange correlation  energies,
and $v_{\rm xc}$ is the exchange and correlation potential\cite{Vxcdens}.
$ \bra H_U \ket$ and $E_{\rm DC}$ depend only on $G^{\rm loc}$ and
are computed from the resolution of the impurity problem. 
With the HI solver, $ \bra H_U \ket$ is
computed with the Migdal formula with a scheme similar
to the one detailed in Ref. \cite{Pourovskii07}.
$E_{\rm DFT}[n({\vr})]$ is computed from the knowledge of the density computed in Eq. \ref{eq:rhoscf}
(see also \ref{app:etot_paw}).
The full localized limit (FLL) version of the double counting is used here as in Ref.\cite{Pourovskii07}.
The application of this double counting with the Hubbard I approximation is described in \ref{app:HubbardI}.

Thus we have the following expression for the total energy in DFT+DMFT:
\begin{equation}
E^{\rm 1}_{\rm DFT+DMFT}= \sum_{\nu,\vk} f^{\rm DFT+DMFT}_{\nu,\vk} \epsilon^{\rm KS-DFT}_{\nu,\vk} + \underbrace{E_{\rm DFT\; DC}[n(\vr)]}_{=-E_{\rm Ha}[n(\vr)]+E_{\rm xc}[n(\vr)]-\int v_{\rm xc} n(\vr) d\vr} + \bra H_U \ket -E_{\rm DC} 
\label{Eq:ldadmftE1}
\end{equation}
Alternatively, one can write:
\begin{equation}
E^{\rm 2}_{\rm DFT+DMFT}= T_0^{\rm DFT+DMFT}  + E_{\rm xc+Ha}[n(\vr)]+\int d\vr v_{\rm ext}(\vr) n(\vr)  + \bra H_U \ket -E_{\rm DC}
\label{Eq:ldadmftE2}
\end{equation}

With $T_0^{\rm DFT+DMFT}$  the analogue of the DFT kinetic energy  in the DFT+DMFT scheme.
The comparison of  \ref{Eq:ldadmftE2} and \ref{eq:edftdmft} yields:
\begin{eqnarray}
T_0^{\rm DFT+DMFT}&=&  \Tr[H_{\rm KS-DFT}G_{\rm DFT+DMFT}] - 2 E_{\rm Ha}[n(\vr)] - \int v_{\rm xc} n(\vr) d\vr -\int v_{\rm ext} n(\vr)d\vr \\
&=&-\int d\vr \sum_{\nu,\nu',\vk} f^{\rm DFT+DMFT}_{\nu,\nu',\vk} \Psi_{\nu,\vk}(\vr)\frac{\nabla^2}{2} \Psi_{\nu',\vk}(\vr)\\
&\equiv&\sum_{\nu,\nu',\vk} f^{\rm DFT+DMFT}_{\nu,\nu',\vk} t_{\nu,\nu',\vk}
\end{eqnarray}
Eq. \ref{Eq:ldadmftE1} is the so called double counting expression of total energy and Eq. \ref{Eq:ldadmftE2} is the direct one.
We emphasize the importance of non-diagonal terms
in the calculation of the true kinetic energy in expression \ref{Eq:ldadmftE2}. This is essential
for the correct calculation of energy.
The exact numerical equality of these two expressions is used in our calculations, both as a test of correctness of the numerical scheme and of the convergence\cite{Torrent08}. This is an important test of the implementation.

This implementation has been done in the code ABINIT\cite{Abinit09} using
the PW-PAW scheme\cite{Torrent08}.

\section{Applications}
\label{sec:appl}
We show here the application of the 
self-consistent DFT+DMFT implementation to Cerium, Ce$_2$O$_3$ and Pu$_2$O$_3$. The first two
systems have been already
studied with a fully self-consistent LDA(ASA)+DMFT technique\cite{Pourovskii07}.
This implementation was based on an LMTO-ASA\cite{andersen_lmto_1975_prb} framework.
We will thus be able both to check if physical effects due to self-consistency are recovered 
in our implementation and what are the improvement brought by the use of the
PW-PAW method.
Indeed, the PW-PAW implementation is more
flexible and more precise: the inclusion of semi-core states is easier
and structural and thermodynamical properties are expected to be more precisely computed.

We have performed both DFT+U calculations and DFT+DMFT calculations.
Moreover for DFT+DMFT,
two solvers were used: the first one -- the static Hartree Fock solver -- , is used in order to check that DFT+U results are recovered with this approach 
(see \ref{subapp:ldautest}) and thus to check a part
of the implementation of the self-consistency over electronic density.
The second solver we used is 
the Hubbard I solver, as in Ref. \cite{Pourovskii07} (see appendix \ref{app:HubbardI}).

For simplicity, we restrict our DFT+DMFT calculations
to the paramagnetic phase of these compounds.

\subsection{ Cerium}
\label{sec:cerium}
Cerium metal exhibits an isostructural phase transition
between the large volume ($\gamma$) phase and a small volume ($\alpha$) phase.
In the $\gamma$ phase, electrons are localized, whereas upon pressure, electrons
and more and more delocalized\cite{johansson74}. Several studies using a variety of methods
have been carried out to understand the electronic structure of the $\gamma$ phase
or the mechanism of the transition.
\cite{Svane96,Laegsgaard99,Shick00,Zolfl01,held_cerium_2001_prl,held_cerium_2003_prb,McMahan05,haule_cerium_prl_2005,Sakai2005,luders05,Amadon06,Amadon08a}
In this DFT+DMFT study, we are using a simplified solver
for the impurity model. We thus study only the $\gamma$ phase
in order to compare our results to the calculations of
Pourovski {\it et al} \cite{Pourovskii07}. We focus
here on the spectral function and on structural properties.

\begin{figure}
{\resizebox{8.0cm}{!}
{\rotatebox{0}{\includegraphics{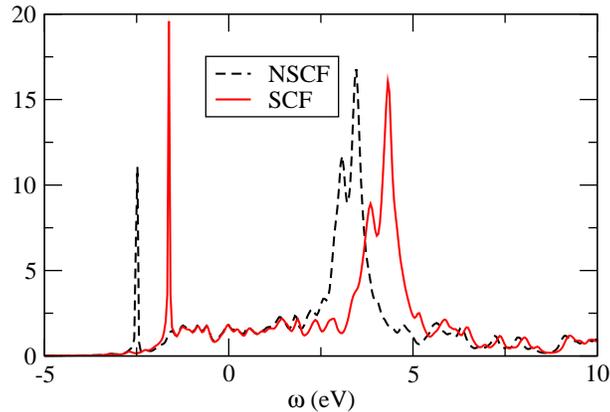}}}}
\centering
\caption{Spectral function of $\gamma$-cerium, computed in LDA+DMFT (HI) with and without
Self-consistency on the charge density.}
\label{fig:cespectral}
\end{figure}

\begin{figure}
{\resizebox{8.0cm}{!}
{\rotatebox{0}{\includegraphics{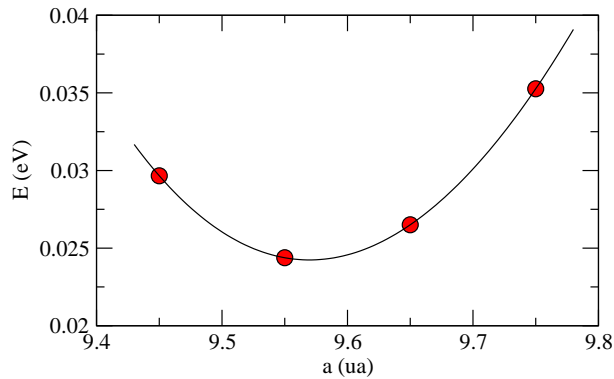}}}}
\centering
\caption{Energy versus volume curve for $\gamma$ Cerium computed in the LDA+DMFT
framework with the Hubbard I solver. As discussed in the text, this energy variation
comes mainly from $E_{\rm LDA}[n_{\rm LDA+DMFT}]$.}
\label{fig:ceev}
\end{figure}

\subsubsection{Computational scheme and details.}

In the PAW datasets, 5s,5p,6s,5d and 4f states
are taken into the valence. Atomic data for PAW are taken from Ref. \cite{Amadon08a}.
The cutoff energy for the plane wave expansion is 12 Ha.
We found that having 28 k-points in the Irreducible Brillouin Zone is
sufficient to have a precision of 0.02 au on lattice parameter and 0.2GPa on bulk modulus.
Calculation of the spectral function is carried out at the experimental volume
(34.8 \AA$^3$). 
All calculations are carried at 273K.

For the DMFT loop, 20 KS functions are used to define correlated orbitals.
They constitute the basis for the DMFT loop. Such as choice is
especially important and should be appropriate to do a
comparison with ASA calculations. This is discussed in 
\ref{app:bands}.

\subsubsection{Spectral functions and structural parameters: results and discussion.}

Spectral functions are gathered on Fig. \ref{fig:cespectral}, and structural
properties are gathered in Tab. \ref{tab:cestruc}.

\begin{table}[h]
\centering
\begin{tabular}{lcc}
\\\hline\hline\hline
 &                                  a (a.u.) & $B_0$ (GPa)    \\
\hline
    Exp\cite{Jeong04,Olsen85}      &    9.76 & 19/21          \\
    PAW/LDA+U                            &    9.58 &  32            \\ 
    PAW/LDA+DMFT NSCF      (H-I)   &    9.41 &  38      \\ 
    PAW/LDA+DMFT SCF       (H-I)   &    9.58 &  36      \\ 
    ASA/LDA+U \cite{Pourovskii07}  &    9.44 &  49        \\
    ASA/LDA+DMFT NSCF (H-I)\cite{Pourovskii07}  &    9.28 &  50        \\
    ASA/LDA+DMFT  SCF (H-I)\cite{Pourovskii07}  &    9.31 &  48   \\
\hline\hline\hline
\end{tabular}
\caption{Lattice parameter $a$ and Bulk modulus $B_0$ of $\gamma$ Cerium according to
experimental data and calculations within LDA+U and LDA+DMFT, in PAW and in ASA\cite{Pourovskii07}.
For a appropriate comparison, PAW/LDA+U are carried out with the expression of the density matrix shown in appendix
\ref{app:compaldau}.}
\label{tab:cestruc}
\end{table}

First, the effect of self-consistency over density
shifts the position of both the lower and upper Hubbard band 
towards higher energies by 0.7 eV.
In comparison the shift was 0.2 eV in the ASA calculation\cite{Pourovskii07}.
Concerning structural properties, the inclusion of self-consistency induces a change of 2\% 
on the lattice parameters, whereas in ASA, the change is only 0.5 \%. 
So the magnitude of the effect of self-consistency is different in LMTO-ASA and in PW-PAW.
Effect of self-consistency over spectra and structural properties
are thus qualitatively similar to the study of Pourovskii {\it et al} but the magnitude of the effect 
is larger in PW-PAW. We emphasize that, it our framework, including semicore states in the valence  is 
easy and natural.
On the contrary, in our ASA calculations of Ref. \cite{Pourovskii07}, the calculation
neglects the 5s and 6p orbitals for the calculation of total energy, and the 5s and 5p for spectra.
These assumptions are physically motivated but restrict the generality of the method.

We then compare ASA and PAW results with experiment.
In our calculations with the parameters used, the  agreement of both spectra and structural properties
with experiment\cite{Wielickka82,Wuilloud83} is slightly better compared
to ASA calculations\cite{Pourovskii07}. 

\subsubsection{Origin of the variation of the energy versus volume.}

We discuss now the behaviour of the energy versus volume curve in DFT+DMFT. We can split
the energy in two contributions: the first one include the interaction U, namely $\bra H_U \ket -E_{\rm DC}$
and the second include all the other terms and is $E_{\rm LDA}[n_{\rm LDA+DMFT}]$.
If we separate these two contributions as a function of volume, one thus see that
the first term is negligible for Cerium, in DFT+DMFT (Hubbard I). 
It comes from the fact that the number of f-electrons found in the solver
is approximately one in this approximation, for each volume.
In other words, the
variation of energy as a function of volume is completely described by $E_{\rm LDA}[n_{\rm LDA+DMFT}]$.
It means that, for each volume, DFT+DMFT converges to a new density, $n_{\rm LDA+DMFT}$ and that  the LDA energy alone
for this set of densities show the behaviour plotted on Fig.\ref{fig:ceev}.

In LDA+U, the energy decomposition show a slightly different behaviour: the term
$\bra H_U \ket -E_{\rm DC}$ is not completely negligible. In fact, computing the energy
variation with only  $E_{\rm LDA}[n_{\rm LDA+U}]$ leads to a lattice
parameter of 9.35 au (about 2\% less than the LDA+U value). It comes directly
from the fact that the number of correlated electrons in LDA+U changes as the volume changes
and modify $\bra H_U \ket -E_{\rm DC}$.
However, the main effect is the same in LDA+U and LDA+DMFT: the change of the electronic density
induced by correlation, shows a variation as a function of volume, which shifts the equilibrium volume
mainly because of the variation of $E_{\rm LDA}[n_{\rm LDA+DMFT}]$.

Let's now analyse the variation of the converged densities as a function of volume with respect to the energy.
In fact, we observe that the variation of $E_{\rm LDA}[n_{\rm}]$ for a given density, can be qualitatively 
linked to the number of
f-electrons: for an increase
of volume of 12\% around the equilibrium volume, the number of $f$ electrons increases of 0.007 in LDA,
and decreases of 0.002 in LDA+U and of 0.03 in LDA+DMFT .
The Hartree interaction energy between $f$ electrons should thus decrease as a function of volume
in the case of LDA+U/LDA+DMFT because the number of $f$ electrons decreases.
In all these three methods, however the impact of these variations are
not the same on the interaction energies: LDA+U and LDA+DMFT give the same equilibrium volume whereas the variations
of the number of electron are very different.
If might come from the different
density matrices which are the results of these three calculations:
In LDA+DMFT, electrons are more homogeneous distributed because of possible
fluctuations between different localized atomic states. On the contrary, in LDA+U, the electron is localized
in a single orbital\cite{LDAnf}.
A general principle is however that Hartree interactions between electrons belonging to different orbitals
are not the same and should be lower than the Hartree self-interaction of one electron in one orbital.
So as Hartree interaction should be lower in LDA+DMFT with respect to LDA+U. It might appear logical
that a large variation of the number of electrons in DFT+DMFT is in fact expected to recover the same variation
of volume.
As a consequence, the variation of the number of electron
as a function a volume is only a qualitative hint to understand the variation of energy, but a complete
understanding requires the knowledge of the density matrix, or of the total electronic density.

In the case of non-self consistent LDA+DMFT calculations, the expression
\ref{Eq:ldadmftE1} is used for the total energy\cite{Amadon06}. As a consequence, the main
effect comes from the band energy. Moreover, one could see that the difference between the DMFT band energy
and the LDA band energy decreases when the volume increases as previously observed\cite{Amadon06}.
However, the intensity of this variation is weaker than the correct variation
at the self-consistent density. Moreover in the SCF calculation, this large variation
is partially compensated by LDA double counting terms. It show that to have a physical
correct behaviour of the different terms of the energy, achieving convergency to the LDA+DMFT
density is mandatory. As emphasized before, this is the very change of the electronic density which
make the difference in $E_{\rm LDA}$. Such change cannot be reproduced
by a non-self-consistent calculation.

\subsubsection{Search for the ground state of the system, metastable states.}

As underlined in the previous paragraph, the orbital
anisotropy obtained in the ground state with LDA+DMFT is very small and is thus different
from the ground state obtained in LDA+U which has a large orbital anisotropy\cite{Shick00,Amadon08a}.
We thus expect the convergency in the case of complex system with large unit cell
(such as $\beta$ cerium) to be much easier with the LDA+DMFT method
than with the LDA+U method\cite{Amadon08a}.

\subsection{ Ce$_2$O$_3$}

Ce$_2$O$_3$ is a Mott antiferromagnetic insulator with an optical band gap of 2.4 eV\cite{Prokofiev96}
and a N\'eel temperature of 9K.
A ionic counting of electron leads to one $f$ electron in the valence band.
The qualitative picture of an insulator has been obtained by first principles calculation 
using the LDA+U\cite{fabris05,ander07,losch07,silva07}, Hybrid functional\cite{silva07}, 
DFT+DMFT framework\cite{Pourovskii07,Jacob08}, and GW+LDA+U\cite{Jiang09}.
In most all these calculations, one electron
is localized and the ground state is an antiferromagnet. 
LDA+DMFT is able to describe a paramagnetic insulating solution\cite{Pourovskii07}. 
In this section, results of the application of the LDA+DMFT scheme to this compound using the PW-PAW framework are given.

\subsubsection{Computational scheme and details.}

PAW atomic data for Cerium are the same as in the preceding section.
For oxygen, 2s and 2p electrons are taken in the valence. PAW matching radius for oxygen is 1.52 ua.
The cutoff energy for the plane wave expansion is 30 Ha. 
We use 32 k-points in the Irreducible Brillouin Zone.
Structural properties are converged to less than 0.01 au for
the lattice parameter and less that 0.1 Mbar for the bulk modulus. 
For simplicity, we do not relax the internal parameters: the volume is the only variable of the system.

Calculations are carried out at a temperature of 273 K.
52 Kohn Sham bands are used to define correlated orbitals and thus as a basis for DMFT calculations.
This number of bands is appropriate to do a comparison with previous 
calculations with the ASA formalism using a basis containing the same number of bands.
The value of U is taken to be 6.0 eV and J is 0 eV.
\begin{table}[h]
\centering
\begin{tabular}{lccc}
\\\hline\hline\hline
                                              &    a (\AA) & $B_0$ (Mbar) \\
\hline
    Exp\cite{Barnighausen1985}                 &    3.89  & 1.11     \\
    PAW/LDA+U(AFM)\cite{silva07}              &    3.87  & 1.3      \\ 
    PAW/LDA+U(AFM)                            &    3.85  & 1.5      \\   
    PAW/LDA+DMFT (H-I) NSCF                   &    3.76  & 1.7      \\   
    PAW/LDA+DMFT (H-I) SCF                    &    3.83  & 1.6      \\     
    ASA/LDA +U \cite{Pourovskii07}            &    3.84  & 1.5      \\         
    ASA/LDA+DMFT(H-I) NSCF\cite{Pourovskii07} &    3.79  & 1.6      \\     
    ASA/LDA+DMFT(H-I) SCF\cite{Pourovskii07}  &    3.81  & 1.6           
\\\hline\hline\hline
\end{tabular}
\caption{Lattice parameter $a$ and Bulk modulus $B_0$ of Ce$_2$O$_3$ according to
experimental data and calculations with various frameworks. In our calculation
and in Ref. \cite{Pourovskii07}, the c/a ratio is fixed to the experimental value
(1.56).  PAW/LDA+U are carried out with the expression of the density matrix corresponding 
to Eq.\ref{eq:dmatpuopt3} and with U=6 eV. Calculations are done at a temperature of 273K.
Note that entropy is neglected in these calculations. Calculations from Da Silva\cite{silva07} and
Pourovskii {\it et al}\cite{Pourovskii07} use respectively U=5.3 eV and 5.4 eV.}
\label{tab:ce2o3compalda}
\end{table}

\subsubsection{Results and discussion.}

\begin{figure}
{\resizebox{8.0cm}{!}
{\rotatebox{0}{\includegraphics{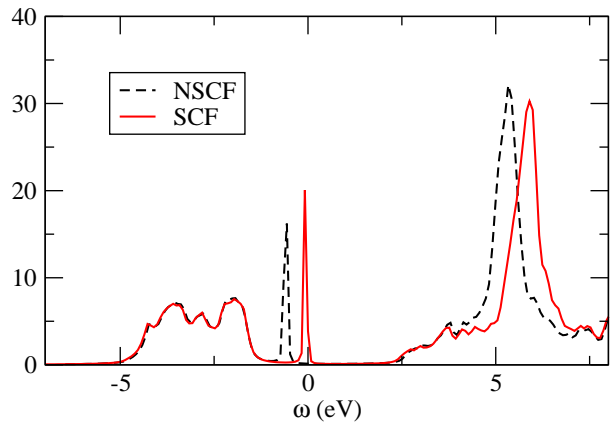}}}}
\centering
\caption{Spectral function of Ce$_2$O$_3$, in LDA+DMFT (Hubbard I) with and without self-consistency.}
\label{fig:ce2o3spectra}
\end{figure}

Spectral functions  are given on Fig. \ref{fig:ce2o3spectra} and
structural properties are gathered on Tab. \ref{tab:ce2o3compalda}.

Concerning first LDA+U results for the structural parameters,
our calculations for the ground state, with antiferromagnetism (AFM) 
are in good agreement with calculations of Ref. \cite{silva07}.
The slight disagreement concerning the bulk modulus might come from the
absence of relaxation of internal parameters in our calculations.

Results from LDA+DMFT calculations reveal that effect of self-consistency over spectra and structural properties
are qualitatively similar to the study of Pourovskii {\it et al}: the volume increases and the gap slightly decreases
in the self-consistent calculation with respect to the non-self-consistent one.
The quantitative results are however different from Ref. \cite{Pourovskii07}:
the ASA approximation induces an slight error of 1\% in the determination of the lattice
parameter with respect to full potential codes. We correct this error in the  PW-PAW implementation of LDA+DMFT calculations.

Spectral function computed with LDA+DMFT with the PW-PAW basis
shows a good agreement with the optical gap of 2.4 eV\cite{Golub95} as encountered also in
LDA+U\cite{silva07}.  However, the oxygen p band is not in good agreement with respect to photoemission spectrum\cite{Allen85}.
GW or hybrid functionals in combination with DFT+U or DMFT can improve the agreement\cite{Jiang09,Jacob08}

\subsubsection{Comparison of LDA, LDA+U and LDA+DMFT densities.}

The analysis of the variation of different energy terms
as a function of volume leads to the same conclusion as for Cerium:
the variation of energy is mainly due to $E_{\rm LDA}$ and is correlated
to the variation of the number of $f$ electron. As the volume increases, 
this number increases in LDA, decreases in LDA+DMFT and is nearly constant in LDA+U.
However, a more detailed view on electronic densities is most useful to illustrate the impact of the different approximations.
In particular, as observed in the case of cerium, the orbital 
anisotropies are different in the LDA+U results and the DMFT calculation.
In particular, one f-orbital is filled with the LDA+U method whereas in DMFT, electrons are allowed
to fluctuate between different orbitals, hence the diagonal terms in the occupation matrix of the
local orbital basis  are similar and not far from 0.07.

\begin{figure}
\centering
\begin{tabular}{ll}
{\resizebox{7.0cm}{!}
{\rotatebox{0}{\includegraphics{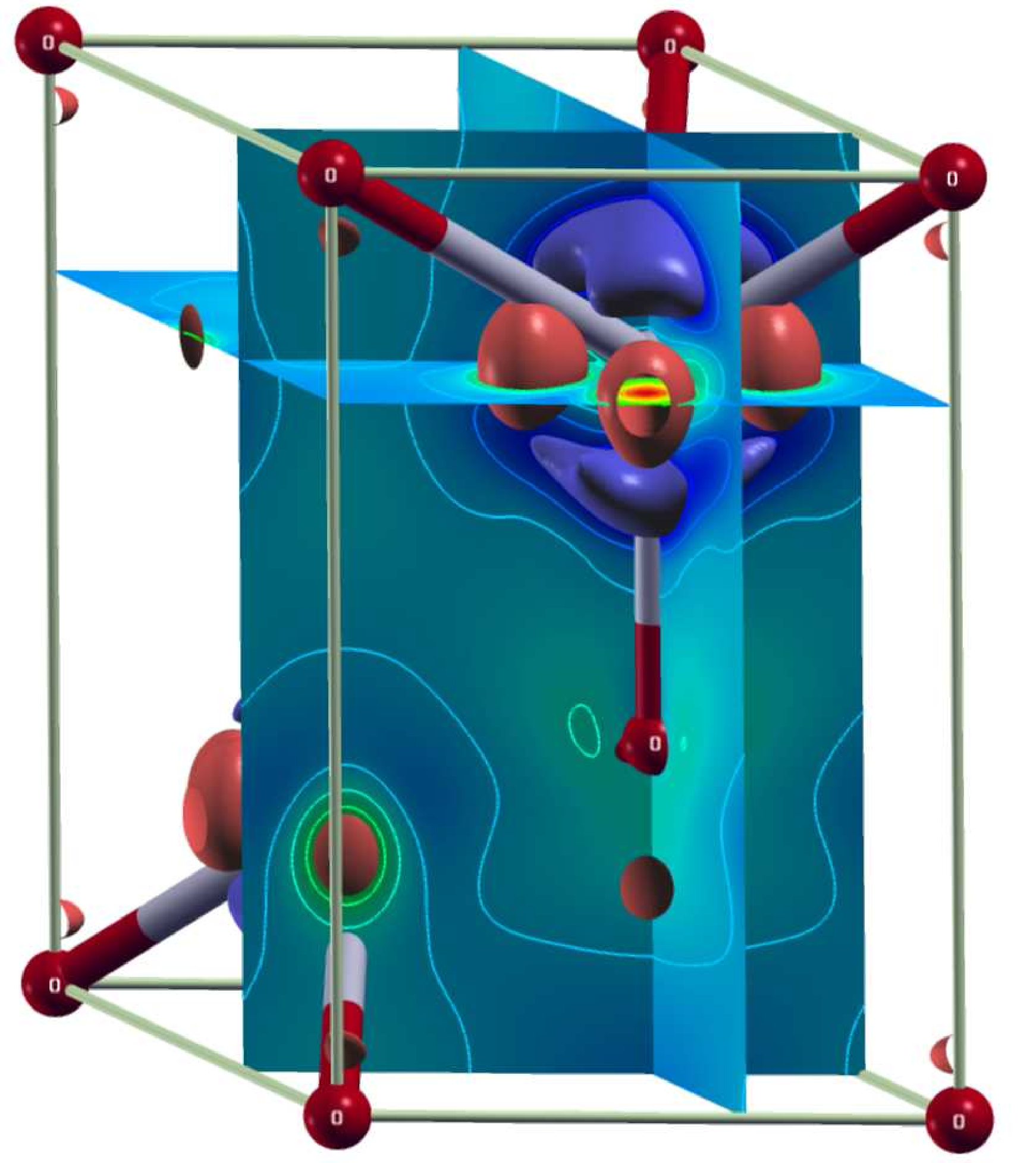}}}}
{\resizebox{7.0cm}{!}
{\rotatebox{0}{\includegraphics{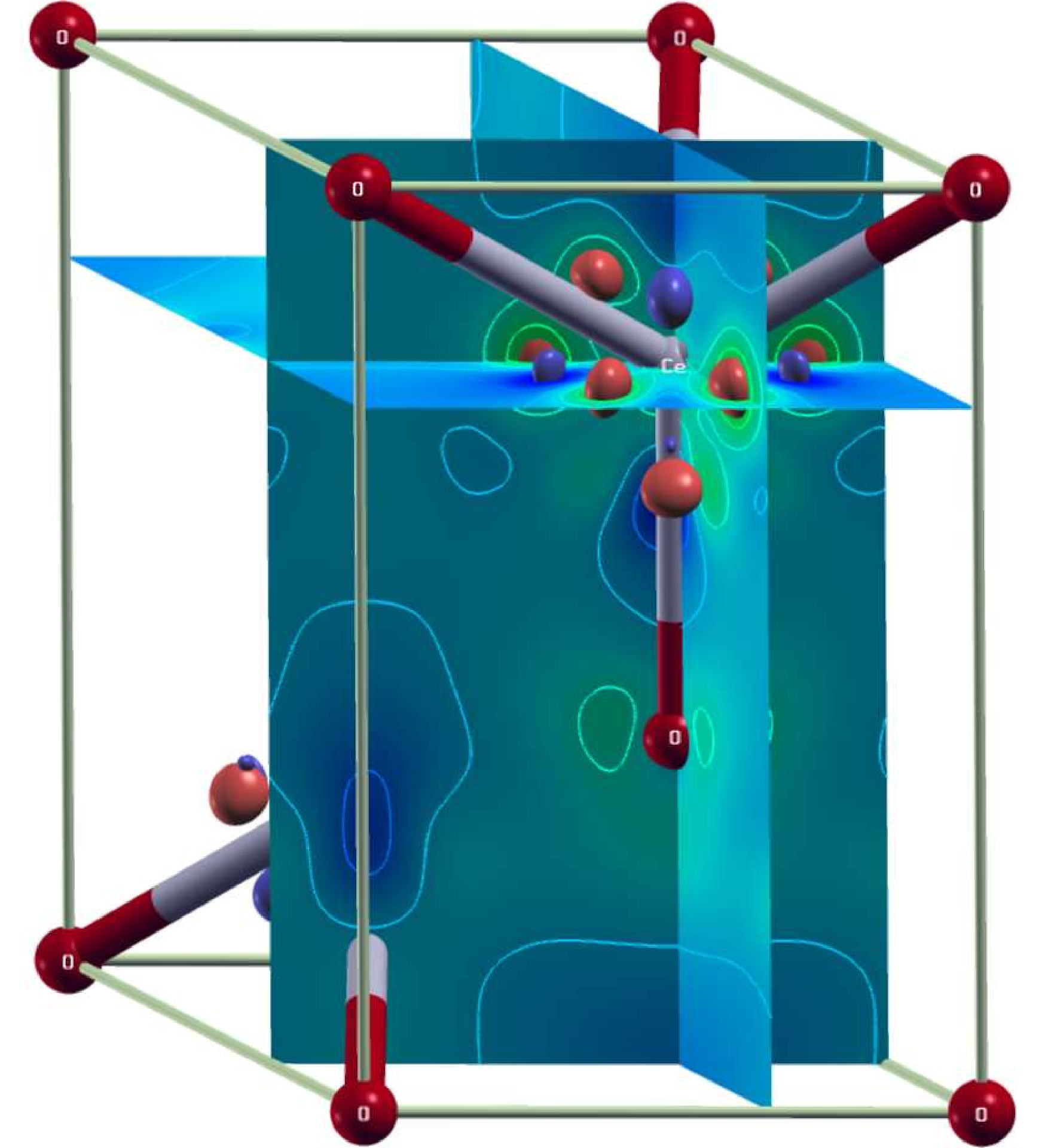}}}}
\end{tabular}
\caption{Difference between electronic densities computed in the LDA+U (left) / LDA+DMFT (right)  approximations 
and in the LDA approximation.
The density of the filled orbital f$^{2-}$ is clearly visible on the left.
Blue (resp. green-red) area corresponds to positive
(resp negative) value of the difference.}
\label{fig:Ce2O3LDAULDADMFT}
\end{figure}

To illustrate the differences between LDA+DMFT and LDA+U electronic densities,
we have plotted\cite{Kokalj03} on Fig.\ref{fig:Ce2O3LDAULDADMFT} 
the isosurfaces of the differences between LDA+U/LDA+DMFT  and LDA electronic densities.
LDA+U shows a localization of one electron in only one f-orbital which is coherent with the breaking
of orbital and spin symmetry in DFT+U.
On contrary, in LDA+DMFT, as a consequence of fluctuations, the f-electron is spread among f-orbitals. 
As a consequence, the difference between LDA+DMFT and LDA densities is weaker.

\subsubsection{Thermodynamical study.}

From the energies obtained for cerium and cerium oxide, one can compute the variation
of internal energy $\Delta_r U$ of the following reaction:
\begin{equation}
2 {\rm Ce} + \frac{3}{2} {\rm O}_2 \rightarrow {\rm Ce}_2{\rm O}_3
\end{equation}

We have thus minimized the structure of the molecule
 ${\rm O}_2$. The distance is found to be 1.22 \AA$^3$ and the cohesive
 energy is 7.6 eV in agreement with previous results (see e.g. Ref. \cite{Jomard08}).

$\Delta_r U$ computed in LDA+DMFT with the Hubbard I solver gives thus a value
of -19 eV $\pm$ 0.1 eV, within 1. eV of the experimental results (-18.63 eV) \cite{Lide1999} and in the range of existing value
computed in LDA, and LDA+U\cite{silva07,ander07,losch07}.
Upon variation of U between 5 and 7 eV, the computed energy show a change of less than 0.15 eV.
This results shows that LDA+DMFT may have the same precision as LDA+U in the calculation
of thermodynamical quantities.  Moreover, LDA+DMFT enables us to compute this energy of reaction
for the true paramagnetic phases of Cerium and ${\rm Ce}_2{\rm O}_3$.

\subsection{Pu$_2$O$_3$}

Pu$_2$O$_3$ is an paramagnetic metal insulator above 10 K\cite{mccart81}.
Its conductivity gap is 1.8 eV and in a ionic picture, it should contain 5 $f$ electrons. 
In this section, we describe this compound with GGA(PBE)+DMFT (Hubbard I) and compute
the spectral properties of this system and we discuss the improvement
with respect to GGA(PBE)+U\cite{Sun08,Jomard08}.

\subsubsection{Computational scheme and details.}

We use the same parameters as for cerium oxide.
For the PAW basis, semicore states are included in the atomic dataset\cite{Jomard08}
and for the PW basis, the cutoff energy for the plane wave is 30 Ha.
The value of U is taken to be 4 eV.
As for cerium oxide, 52 Kohn Sham bands are used to compute Wannier $f$ orbitals.

\subsubsection{Results and discussion.}

\begin{figure}
{\resizebox{8.0cm}{!}
{\rotatebox{0}{\includegraphics{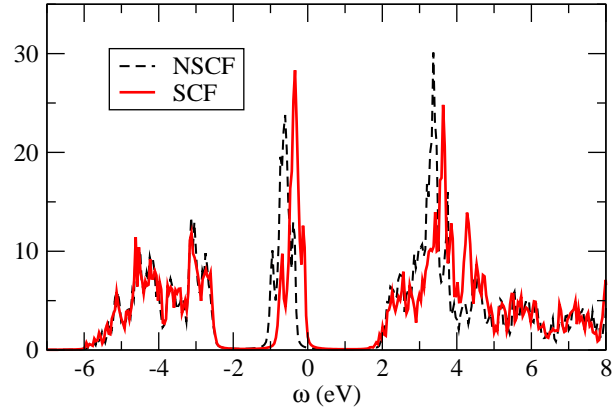}}}}
\centering
\caption{Spectral function of Pu$_2$O$_3$, in GGA+DMFT (Hubbard I) with and without self-consistency.}
\label{fig:pu2o3spectra}
\end{figure}

GGA+DMFT spectral functions obtained from the SCF and NSCF calculations are shown on Fig \ref{fig:pu2o3spectra}.
The GGA+DMFT calculations describes Pu$_2$O$_3$ as a Mott Hubbard insulator with
a f-f gap of ~1.8 eV. This is similar to what is obtained with the GGA+U method\cite{Jomard08}.
However, in GGA+DMFT, we are able here to describe the paramagnetic phase.
Concerning self-consistency effects, it shifts the f peak towards higher energy with respect
to the O-p band and to the conduction band. This is similar to the behaviour encountered for
Ce$_2$O$_3$.

The agreement with experimental data is good.
The gap is near the experimental conductivity gap of 1.8 eV\cite{McNeilly64} -- which contains,
however, two particles excitations. 
Structural properties, shown on Tab. \ref{tab:pu2o3} are also in rather good agreement with experiment.

As observed in cerium oxide, the density matrix obtained in the paramagnetic solution 
highlight particularly the role of fluctuations inside the atomic orbitals
brought by the Hubbard I solver.
However, contrary to the case of Ce$_2$O$_3$, there is
still an orbital anisotropy coming mainly from crystal field :
crystal field splitting is indeed larger in plutonium oxide,
in agreement with the fact that Cerium $f$ orbitals
are more localized than Pu $f$ orbitals.
From the energy levels computed in Hubbard I (expression \ref{eq:levels}),
we can estimate the crystal field splitting to 0.05 eV in  Ce$_2$O$_3$ and
0.5 eV in  Pu$_2$O$_3$.  
It thus explains why orbital anisotropy is
observed in  Pu$_2$O$_3$ and not in  Ce$_2$O$_3$: it comes from the direct physical
effect that lower levels have more statistical weights and in  Pu$_2$O$_3$, the difference
in energy between levels is sufficient.
Indeed and logically, it is the same orbitals that are filled in LDA+U\cite{Jomard08} that are partially filled in LDA+DMFT.
Nevertheless, in LDA+U, one can encounter
spurious orbital anisotropy and associated metastables states that are due
to a breaking of symmetry, or to the filling of higher states in energy and the subsequent stabilization
of their energy levels. 
As observed in Ref. \cite{Pourovskii09}, such breaking of symmetry does not happen
in GGA+DMFT because this method takes into account all possible Slater determinant in
the Green's function: metastables states encountered in LDA+U which have a nearly degenerate energy
are treated on the same footing in the DMFT Green's function.

\begin{table}[h]
\centering
\begin{tabular}{lccc}
\\\hline\hline\hline
                                                & V (\AA$^3$) & $B_0$ (GPa) \\
\hline
    Exp\cite{Wulff1988,McCart1981}              & 75.5 -76.1  &             \\
    PAW/GGA+U(AFM)\cite{Jomard08}               &  78.1       & 137         \\ 
    PAW/GGA  (AFM)\cite{Jomard08}               &  73.4       & 142         \\   
    PAW/GGA+DMFT (H-I) NSCF                     &  74.5       & 143         \\   
    PAW/GGA+DMFT (H-I) SCF                      &  75.7       & 139         
\\\hline\hline\hline
\end{tabular}
\caption{Lattice parameter $a$ and Bulk modulus $B_0$ of Pu$_2$O$_3$ according to
experimental data and calculations with various frameworks. In our calculation
the c/a ratio is fixed to the experimental value (1.542\cite{Wulff1988}) and are done at a temperature of 273K.
Note that entropy is neglected in these calculations. Calculations from Jomard\cite{Jomard08} 
uses U=4.0 eV and 0.7 eV, whereas we use U=4.0 eV and J=0 eV.}
\label{tab:pu2o3}
\end{table}

\section{Conclusion}

In this paper, a new implementation of self-consistent DFT+DMFT and
application of this formalism to some f-electrons systems are presented.

From the methodological side, a self-consistent implementation of DFT+DMFT
using the projected local orbital formulation is shown.
This implementation is done inside a plane wave-projector augmented wave (PAW) framework.
It allows to carry out DFT+DMFT calculations of spectra and total energy with the precision of an all-electron code
and the flexibility of a plane wave code, which opens the way for relaxation and dynamics.
The applications are carried out with the Hubbard one solver, but the scheme is completely general
and could be used in combination with Quantum Monte Carlo Solver.
Moreover we discuss the comparison of this scheme with schemes using 
MLWF or ASA basis and explain how to compare these schemes for f-electrons systems.
We stress that for f-electrons systems, DFT+DMFT is particularly adapted but that such systems
are notoriously difficult to treat using MLWF in combination with DFT+DMFT. The current framework, which avoids
the construction of MLWF is thus well adapted and we show its success for several f-electrons compounds.

We show the application of this framework to three strongly correlated systems with $f$ orbitals.
First, we study cerium and  Ce$_2$O$_3$. 
We compare our calculations to a previous study using the ASA basis.
This is the opportunity to highlight the improvement brought
by a correct description of the potential and the density  with respect to ASA:
structural parameters are largely improved and thermodynamical quantities, such as formation energies, can be computed.
More generally, with this new implementation,
the calculation of energy in DFT+DMFT is on the same footing as the same calculation
in LDA: semicore states are easily included, energy of different structures can be easily compared.

Finally, we show the first study of structural and spectral properties of plutonium sesquioxide within DFT+DMFT. 
We compare this compound to cerium sesquioxide and underline the
reduced occurrence of spurious metastable states in DFT+DMFT in comparison to DFT+U.

The existence of non-physical metastable states is an important drawback of DFT+U which prevents to
find the ground state of large heterogeneous systems, especially for $f$ elements.
\cite{Shick00,Larson07,Amadon08a,Jomard08,Dorado09,Jollet09}. 
Therefore, DFT+DMFT, besides giving a better physical description of correlated
systems, suppresses at least partially this problem.
The computational cost of DFT+DMFT is however much larger than DFT+U but could be
much reduced by using a simplified solver, such as Hubbard I,
provided the system studied is a Mott insulator.

The general DFT+DMFT framework with the 
accurate PW-PAW basis opens the way to the calculations of
others properties, such as forces, phonons\cite{Audouze06}.
Our implementation has been done inside the open source code ABINIT\cite{Abinit09,abinit3}.

Note Added: After the writing on the manuscript was completed, we became aware of a related work of
J.Z.Zhao {\it et al} (arXiv:1111.2157v2) on an implementation of LDA+DMFT in a pseudopotential plane wave framework.

\ack
I thank M. Torrent and F. Jollet for fruitful discussions.
I am also grateful to Yann Pouillon for helpful technical support (with the build system of ABINIT).
This work was granted access to the HPC resources of CCRT under the allocations
2010096008 and 2011096681 made by GENCI (Grand Equipement National de Calcul Intensif). We thank the CCRT team for support.
The work was supported by ANR under project CORRELMAT.

\appendix
\section{Self-consistency over charge density: peculiarities of the PAW formalism}
\label{app:scpaw}

In this appendix, we derive the expression for the density used
in self-consistent DFT+DMFT within PAW.

The electronic density (we neglect spin in this section)
is obtained from \ref{eq:rhoscf}. We have
\begin{equation}
\label{eq:dstnd}
n({\bf r })= \sum_{\nu,\nu',{\bf k}} f_{\nu,\nu',\vk} \bra \Psi_{\nu,{\bf k}}| {\bf r} \ket \bra {\bf r}  |\Psi_{\nu',{\bf k}} \ket.
\end{equation}

Then, using the fundamental relation (11) of Ref. \cite{Blochl94} for the operator $| {\bf r} \ket \bra {\bf r}  |$,
we have
\begin{eqnarray}
\nonumber
n({\bf r })&=&\sum_{\nu,\nu',{\bf k}} f_{\nu,\nu',\vk}  \tPsi^*_{\nu,{\bf k}}({\bf r})\tPsi_{\nu',{\bf k}}({\bf r}) + \\
\nonumber
&&\sum_{ij}\sum_{\nu,\nu',{\bf k}}  f_{\nu,\nu',\vk}\bra \tPsi_{\nu,{\bf k}}|\tp_i \ket ( \bra  \varphi_i | {\bf r} \ket \bra {\bf r}| \varphi_j\ket - \\
&&\bra  \tphi_i | {\bf r} \ket \bra {\bf r}| \tphi_j\ket) \bra \tp_j |\tPsi_{\nu',{\bf k}} \ket. 
\label{eq:nnd}
\end{eqnarray}
where $\tp_j$ is the $n_j^{\rm th}$ projector, for
the angular momentum $l_j$, and its projection $m_j$ on atom $\bR$. 
$\tphi$ and $\varphi$ are respectively the pseudo and all-electron atomic wavefunctions.
$\tPsi$ and $\Psi$ are respectively the pseudo and all-electron Kohn Sham wavefunctions.
Index for atoms are neglected on $\tp$, $\varphi$ and $\tphi$.
It can be rewritten as:
\begin{equation}
n({\bf r })= \tn'({\bf r})+ n^{1'}({\bf r})-\tn^{1'}({\bf r})
\end{equation}
with
\begin{eqnarray}
 \tn'({\bf r}) &=&\sum_{\nu,\nu',{\bf k}} f_{\nu,\nu',\vk}  \tPsi^*_{\nu,{\bf k}}({\bf r})\tPsi_{\nu',{\bf k}}({\bf r})  \\
 n^{1'}({\bf r}) &=&\sum_{ij} \rho'_{ij} \varphi_i({\bf r }) \varphi_j({\bf r }) \\
 \tn^{1'}({\bf r}) &=&\sum_{ij} \rho'_{ij} \tphi_i({\bf r }) \tphi_j({\bf r })
\end{eqnarray}
and
\begin{equation}
\rho'_{ij}=\sum_{\nu,\nu',{\bf k}} f_{\nu,\nu',\vk} \bra \tPsi_{\nu,{\bf k}}|\tp_j \ket
\bra \tp_i | \tPsi_{\nu',{\bf k}} \ket.
\label{eq:rhoijnd}
\end{equation}

So, for the {\it on site} terms of the density, we just have to compute $\rho'_{ij}$,
and then use it instead of $\rho_{ij}$ everywhere in the formalism.

\section{Definition of local orbitals and comparison of DFT+DMFT and DFT+U implementations}
\label{app:compaldau}

DFT+U and DFT+DMFT can be seen as two different approximations for the same 
functional. 
DFT+U is a static mean field solution, and DFT+DMFT contains dynamical fluctuations.
However, their practical implementations are quite different.
In DMFT, one need a frequency dependant local Green's function whereas in DFT+U,
only the integral of it is used.
In this technical appendix, we first briefly compare the definition of
local orbitals in these two methods
(\ref{subapp:localorb_general})
, then we compare the implementations in the PAW
method
(\ref{subapp:localorb})
, and we give a careful numerical comparison with
our implementations
(\ref{subapp:ldautest}).

\subsection{Definition of local orbital, and density matrix in the PLO scheme for a complete KS basis}
\label{subapp:localorb_general}

We compare here the DFT+DMFT PLO formulation and the DFT+U treatment of local orbitals in a specific case.
In DFT+DMFT, projections in the PLO scheme are defined by Eq. \ref{eq:pro_chipsi}.
We assume here that $\chi^{\bR}_{m}$ are a set of
orthonormalized orbitals, such as MLWF or atomic orbitals.
If we also use an infinite window for the construction of $\tchikm$ (Eq.\ref{eq:tchi}) 
then we have $\tchikm=|\chi^{\bR}_{\vk,m}\ket$, because the basis of Kohn Sham 
functions is complete.
In this case, the construction of Wannier orbitals
from  $\chi^{\bR}_{m}$  is just a renormalization with $\bra \chi^{\bR}_{m} |\chi^{\bR}_{m}\ket= 
\bra \chi^{\bR} |\chi^{\bR}\ket$.

In this last case, the density matrix of correlated electrons can be written as
\begin{eqnarray}
n_{m,m'}^{\vR,\sigma}&=&\sum_{\vk\nu} f_{\vk,\nu}^{\sigma} 
\frac{\bra \chi_{m}^{\vR} | \Psi^\sigma_{\vk,\nu} \ket \bra  \Psi^\sigma_{\vk,\nu} | \chi_{m'}^{\vR} \ket}
{\bra\chi^\vR|\chi^\vR\ket
}
\label{eq:densmat}
\label{eq:dmatpuopt3}
\end{eqnarray}

So, if a DFT+U calculation is carried out with this formulation of the density matrix, results could
be compared to a DFT+DMFT calculation with the PLO scheme at convergence of the KS basis.

\subsection{Application to the PAW formalism}
\label{subapp:localorb}

In our PAW implementation of the PLO scheme for DFT+DMFT, we follow the lines 
of the Ref. \cite{Amadon08}. Equation \ref{eq:pro_chipsi} thus writes as Eq (A.2) of Ref. \cite{Amadon08}
in the PAW method:
\begin{equation}
\label{eq:Ppaw}
P^{\bR}_{m\nu}(\bk) = \sum_{n} 
\bra \chi_m^{\bR} | \varphi^\bR_{n} \ket
\bra \tp^\bR_{n} | \tPsi_{\bk\nu} \ket,
\end{equation}
where $\tp^\bR_n$ is the projector $n$ for
angular momenta $l$, and its projection $m$ on atom $\bR$ and
$\chi_m^{\bR}$ is chosen to be the atomic eigenfunctions for the given
angular momentum.

The $\chi^{\bR}_{m}$ are not completely orthonormalized
because in our implementation, they are used only inside the PAW sphere.
However, 96\% of their density is located inside the sphere, so in this discussion, we neglect
this slight non-orthonormalisation.

As we use only the part of the density matrix computed
inside the sphere with the all-electron wave function (see Ref. \cite{Amadon08,Amadon08a}),
$\bra\chi^\vR|\chi^\vR\ket$  represent the integral of the atomic wavefunction
inside the PAW sphere.
Eq. \ref{eq:densmat} thus writes, in the PAW framework, with  ($m$ index are removed for simplicity in the equation because
integrals are done here over radial part of wavefunction)
\begin{equation}
n_{m,m'}^{\vR,\sigma}=
\sum_{n,n'} \rho_{m,n;m',n'}^{\vR,\sigma} \frac{\bra \varphi^\bR_{n} |\chi^\bR \ket \bra 
\chi^{\bR} | \varphi^\bR_{n'} \ket}{\bra\chi^\vR|\chi^\vR \ket}.
\label{eq:noccmmp3}
\end{equation}
It is thus the density matrix in the subset of correlated orbitals used in DFT+DMFT at convergence
of the KS basis.
We will compare in the next subsection the application of this expression in DFT+U to the PLO
scheme in DFT+DMFT.

Another way of computing the projection \ref{eq:pro_chipsi}
and occupation matrices exists, such as an integrated value
in PAW sphere of angular-momentum-decomposed charge densities\cite{Bengone00,Karolak10}.
The slight drawback of such a scheme is that the starting atomic orbitals $|\chi^\vR\ket$
are not easily defined in this approach. The advantage 
is that the spectral weight is not lost\cite{Haule10}
as long as the projection is done with all terms of Eq. (A.1) of Ref. \cite{Amadon08}.
Differences between this scheme and the latter are however small\cite{Amadon08a,Karolak10}.

We underline that atomic data that are used with this framework
have to be transferable in a large energy range because
unoccupied excited states are used in the framework. In particular,
the closure relation over KS states do
not hold if atomic data are not transferable in the relevant energy range.

\subsection{Numerical comparison of the two schemes}
\label{subapp:ldautest}

\begin{figure}
{\resizebox{8.0cm}{!}
{\rotatebox{0}{\includegraphics{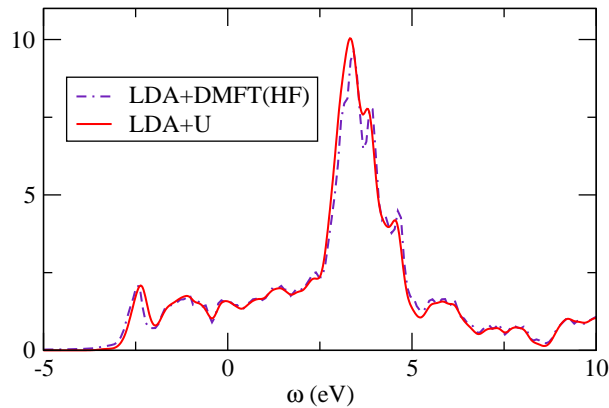}}}}
\centering
\caption{Spectral function of $\gamma$-cerium, computed in DFT+U and DFT+DMFT (HF).}
\label{fig:correl}
\end{figure}

We have carried out this comparison for the volume and spectra of $\gamma$ Cerium.
Parameters for the calculation are similar to those of section \ref{sec:cerium}.
The calculations are carried out at the temperature of 273K. 
Moreover, calculations using DFT+U are done with expression \ref{eq:noccmmp3} 
for the density matrix  and DFT+DMFT uses 30 bands ranging from -32 eV  to  30 eV (Fermi
level is at zero energy).
To assess the lack of completeness of the Kohn Sham basis, one
can compute the  value of $\bra\chi^\vR|\chi^\vR\ket$ using the closure relation:
\begin{equation}
\bra\chi^\vR|\sum_{\vk,\nu\sigma} [ \Psi^\sigma_{\vk,\nu}\ket \bra \Psi^\sigma_{\vk,\nu}] |\chi^\vR\ket.
\end{equation}
For our PAW atomic data, and the bands chosen (see caption of Tab. \ref{tab:celdau}),
we find an average value of 0.95, whereas the numerical value of $\bra\chi^\vR|\chi^\vR\ket$ is
0.964. The Kohn Sham basis thus describes the atomic wavefunction with an
error of less that 2\%.

Spectral functions for the two formulations of DFT+U, as shown on Fig.\ref{fig:correl}, are nearly identical. 
Structural properties, gathered in Tab. \ref{tab:celdau}
show variations smaller than 0.1\% on the lattice parameters and
1\% on the bulk modulus. It is a confirmation of the coherence of the two implementations.

\begin{table}[h]
\centering
\begin{tabular}{lcc}
\\\hline\hline\hline
 &                                  a (a.u.) & $B_0$ (GPa) \\
\hline
    PAW/LDA+U                      &    9.58 &  32        \\ 
    PAW/LDA+DMFT(HF)               &    9.59 &  31        
\\\hline\hline\hline
\end{tabular}
\caption{Lattice parameter $a$ and Bulk modulus $B_0$ of $\gamma$ Cerium according to
calculations using LDA+U with expression \ref{eq:noccmmp3} for the density matrix  and LDA+DMFT using 30 bands ranging from -32 eV  to  30 eV around the Fermi level.}
\label{tab:celdau}
\end{table}

\section{Impact of the energy windows used to define
Wannier functions on physical properties}
\label{app:bands}

In this section, we study the dependency of the spectral function 
as a function of the number of KS states used to define Wannier functions.
The calculations are done in the case of $\gamma$ Cerium.
It is an important issue because the choice of the KS basis implicitly defines the local
orbital subset as an orthonormalized linear combination of these states.

\begin{figure}
{\resizebox{8.0cm}{!}
{\rotatebox{0}{\includegraphics{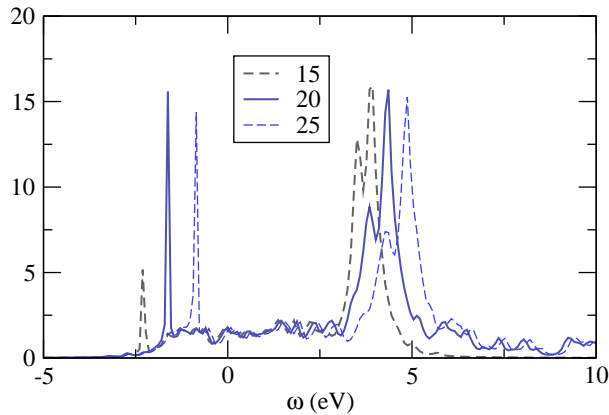}}}}
\centering
\caption{Spectral function of $\gamma$-cerium, computed in LDA+DMFT (HI) for different
choices of the KS basis.}
\label{fig:KSCe}
\end{figure}

Spectral functions for different choices of the KS basis are plotted on Fig.\ref{fig:KSCe}.
As more and more KS states are included, lower (resp. upper) Hubbard band are shifted from ~-2.5 eV to ~-1.eV
(resp. 3.5 eV to 5 eV).
In the Hubbard I approximation used here, the positions of Hubbard bands
are governed by level positions computed in Eq. \ref{eq:levels}.
In this equation, the double counting correction
is computed with the number of electrons used in the Hubbard I solver\cite{Pourovskii07}, which
is nearly one, and does not depend on the window of energy. 
The shift of Hubbard band thus comes 
from the definition of the LDA atomic levels (Eq. \ref{eq:levels}):
As more and more Kohn Sham states are included
in the sum, the mean energy of levels increases tangentially, until
the basis is complete. However we stress that this effect is substantially increased by the self-consistent loop.
The consequence of the higher energy level of $f$-orbitals, is
an increase of the number of $f$ electrons (See Table \ref{tab:KSCe})
because of a larger hybridization with other orbitals.
We stress that in this limit, we are moving away from the atomic limit and 
the Hubbard I solver is not adequate.
Besides, a more extensive physical 
investigation would require the computation of a different value of U for each window of energy (or number
of KS states) used to define the correlated Wannier functions: as the windows
of energy is increased, the orbital are more and more localized and thus the value of U
should increase. 

However, we emphasize that
the results of this convergence only corresponds
to a given choice of correlated orbitals, namely, atomic orbital. Another
legitimate choices exist:
We found, that, in order
to do an adequate comparison between the PLO scheme and a scheme using LMTO-ASA,
it is especially important to use the same number of KS states as the number of muffin tin orbitals
used in the LMTO-ASA calculation. It comes from the fact that Wannier functions which are used as
correlated orbitals in the two schemes are thus built with the same window of energy.
For $\gamma$ Cerium, the ASA calculation of Ref. \cite{Pourovskii07}
are done with a basis set of 20 states (5s, 5p, 6s, 6p, 5d and 4f states).  A calculation
with the PLO scheme must use 20 bands in the KS basis in order to use roughly the same
correlated orbitals.

\begin{table}[h]
\centering
\begin{tabular}{lccc}
\hline\hline\hline
    Number of KS bands             & 15      & 20      & 25      \\
\hline
    Energy range (Ha)              & 1.4     &  1.8    &  2.0    \\ 
\hline
    Number of $f$ electrons        & 1.03    &  1.116  &  1.14     \\ 
\hline\hline\hline
\end{tabular}
\caption{Range of energy for the LDA eigenvalues for several sets of KS states.
The lowest KS states, corresponding to 5s orbitals are located -1.25 Ha under
the Fermi level. Number of electrons obtained in the corresponding LDA+DMFT calculations}
\label{tab:KSCe}
\end{table}

\section{Total energy in DFT+DMFT in the PAW formalism}
\label{app:etot_paw}

In this section, we briefly show that the knowledge of $f_{\nu,\nu',\vk}$
is sufficient to compute all energy terms in the PAW method. We use some
notations of Ref. \cite{Torrent08}.

According to the calculation of $f_{\nu,\nu',\vk}$, one may compute $\rho_{ij}$ and the pseudo-density
with Eq. \ref{eq:nnd} and \ref{eq:rhoijnd}.
All the terms  of the energy which depends on the 
local quantities ($E^1$, $\widetilde{E}^1$.  $E^1_{\rm dc}$, $\widetilde{E}^1_{\rm dc}$) 
are computed with $\rho'_{ij}$, $n^{1'}$ and $\tn^{1'}$ (see Ref. \cite{Torrent08}).
In the plane wave basis, $E_{\rm xc}$, $E_{\rm Ha}$  are  computed directly from the computed DFT+DMFT density.
The sum over eigenvalues and the kinetic energy are directly computed from above
( Eq.\ref{Eq:ldadmftE1} and Eq.\ref{Eq:ldadmftE2} ). We emphasize that there is no approximation to a tight
binding model for the computation of the kinetic energy or the sum
over KS eigenvalues.

\section{Hubbard I solver and interaction}
\label{app:HubbardI}
To solve the impurity model, we use in this paper the Hubbard I method.
This method neglects the hybridization between the impurity
and the bath in the resolution of the Anderson impurity model.
We implement this method
in the same spirit as in Ref. \cite{Pourovskii07}: we impose
that the impurity Green's function and the local Green's function
have the same limit at high frequencies. It follows the following
definition for the atomic level (see also Ref. \cite{Aichhorn09}):
\begin{equation}
\epsilon^{\bR\sigma}_{\rm m,m'}= 
 \sum_{\vk\nu} \bar{P}^{\bR\sigma}_{m\nu}(\bk) \epsilon^\sigma_{\nu \bk} \bar{P}^{\bR\sigma*}_{\nu m'}(\bk) -\Sigma^{\bR,\sigma}_{\rm DC} -\mu.
\label{eq:levels}
\end{equation}

As HI solves an atomic problem without coupling to the bath, the number
of electrons computed in the resolution of the impurity problem is nearly an integer (see Ref. \cite{Pourovskii07}).
For the sake of comparison with Ref.\cite{Pourovskii07}, this number is used to compute the FLL double counting. However we stress that our implementation
is completely general and that the value of the double counting used in Hubbard I is just a consequence
of the approximate nature of the solver HI.

We choose J=0 in all calculations and we precise the value of the interaction U chosen for each system.
As Hubbard I approximation is a valid approximation for well localized systems, we expect however
this approximation to be valid only for large values of U and/or low hybridization (ie high volumes). In other cases,
a more accurate solver should be used.

\clearpage

\begin{thebibliography}{10}
\bibitem{Gloc}
$G^{\rm loc}$ stands for $G_{LL'}^{\rm loc}(i\omega_n)$ where $L$ and $L'$ are
  indices for spin and orbital momenta.
\bibitem{Vxcdens}
$v_{\rm xc}$ depends on the density. This dependance is not vritten in order to
  lighten the equations.
\bibitem{LDAnf}
whereas in LDA, electrons are distributed among orbitals according to crystal
  field.
\end{thebibliography}

\end{document}